\documentclass[showkeys,reprint,
superscriptaddress,
 amsmath,amssymb,
 aps,
prb,
floatfix,]{revtex4-1}
\usepackage{graphicx}
\usepackage{dcolumn}
\usepackage{bm}
\usepackage{epstopdf}

\usepackage{amsmath,amssymb,amsfonts}
\usepackage{fancyhdr}
\usepackage{lipsum}
\usepackage{subcaption}
\usepackage{multirow}
\begin{document}

\author{Soumen Mandal}
\email{mandals2@cardiff.ac.uk}
\affiliation{School of Physics and Astronomy, Cardiff University, Cardiff, UK}
\author{Evan L. H. Thomas}
\affiliation{School of Physics and Astronomy, Cardiff University, Cardiff, UK}
\author{Callum Middleton}
\affiliation{Center for Device Thermography and Reliability, Bristol University, Bristol, UK }
\author{Laia Gines}
\affiliation{School of Physics and Astronomy, Cardiff University, Cardiff, UK}
\author{James Griffiths}
\affiliation{Department of Materials Science and metallurgy, University of Cambridge, Cambridge, UK}
\author{Menno Kappers}
\affiliation{Department of Materials Science and metallurgy, University of Cambridge, Cambridge, UK}
\author{Rachel Oliver}
\affiliation{Department of Materials Science and metallurgy, University of Cambridge, Cambridge, UK}
\author{David J. Wallis}
\affiliation{Department of Materials Science and metallurgy, University of Cambridge, Cambridge, UK}
\affiliation{School of Engineering, Cardiff University, Cardiff, UK}
\author{Lucy E. Goff}
\affiliation{Department of Physics, University of Cambridge, Cambridge, UK}
\author{Stephen A. Lynch}
\affiliation{School of Physics and Astronomy, Cardiff University, Cardiff, UK}
\author{Martin Kuball}
\affiliation{Center for Device Thermography and Reliability, Bristol University, Bristol, UK }
\author{Oliver A. Williams}
\email{williamso@cardiff.ac.uk}
\affiliation{School of Physics and Astronomy, Cardiff University, Cardiff, UK}

\title{Surface zeta potential and diamond seeding on gallium nitride films}

\begin{abstract}
Measurement of zeta potential of Ga and N-face gallium nitride has been carried out as  function of pH. Both the faces show negative zeta potential in the pH range 5.5-9. The Ga face has an isoelectric point at pH 5.5. The N-face shows higher negative zeta potential due to larger concentration of adsorbed oxygen. Zeta potential data clearly showed that H-terminated diamond seed solution at pH 8 will be optimal for the self assembly of a monolayer of diamond nanoparticles on the GaN surface. Subsequent growth of thin diamond films on GaN seeded with H-terminated diamond seeds produced fully coalesced films confirming a seeding density in excess of 10$^{12}$ cm$^{-2}$. This technique removes the requirement for a low thermal conduction seeding layer like silicon nitride on GaN.
\end{abstract}

\maketitle

\section{Introduction}
With the high breakdown voltage and current handling ability, gallium nitride (GaN) high electron mobility transistors (HEMT)  are the current benchmark for high-power, high-frequency applications\cite{mishra2002, mishra2008}. However, the heat generated during high power density conditions limits the use of such devices, with Lee \textit{et al.}\cite{lee2008} demonstrating that small increases in the operating temperature can lead to drastic reductions in device lifetime. To realise long-lifetime and high performance of GaN electronics it is therefore essential to effectively extract  heat from the devices. The current state-of-the-art structures are fabricated from GaN grown on top of SiC substrates\cite{gaska1998, kuball2002}, widely used for RF applications due to its superior thermal conductivity ($\kappa_{SiC} \approx 360-490$ W/m$\cdot$K\cite{goldberg2001}). The use of SiC has allowed good device performance in terms of thermal management, but replacing SiC with diamond substrates of thermal conductivity up to $\kappa_{Diam} \approx 2100$ W/m$\cdot$K\cite{ho1972}, more significant advances in GaN electronic devices will be possible.

{To fabricate GaN on diamond devices, the diamond can either be grown on top of the GaN or vice versa.} {While previous trials of} GaN grown on single crystal diamond  have led to  promising results\cite{jessen2006, hirama2011}, the high cost associated with single crystal diamond makes this technique unattractive.  Alternatively, polycrystalline diamond can be grown on GaN through the use of an adhesion layer \cite{dumka2013, pomeroy2014}.However, while the polycrystalline diamond has excellent thermal properties\cite{coe2000}, the poor thermal  {conductivity} of the typically amorphous adhesion layer {presents a barrier to the extraction of heat from the} devices\cite{sun2015}.  {Therefore}, it is  {crucial} to grow the diamond layer directly {onto the surface of the GaN to provide more effective thermal management}. Initial attempts have been made with mixed success\cite{gu2016}. A further complication for the case of diamond growth on GaN is that wurtzite GaN has two distinct terminations along the polar direction. For (0001) GaN the surface is Ga-terminated and for the (000$\bar{1}$) surface the GaN is N-terminated. In the majority of cases GaN HEMT layers are grown on the Ga- face and therefore the diamond heat spreading layer must be produced on the back N-face. Since the GaN crystal is polar along the [0001] axis it might be expected that the different crystal surfaces could have very different charge states and therefore understanding diamond film nucleation on the appropriate face is vital.

Due to large differences in surface energies between diamond ($\sim 6$ J/m$^2$)\cite{harkins1942} and {the} most commonly used Si ($1.5$ J/m$^2$)\cite{williams2011rev} substrates for diamond growth, attempts at heteroepitaxial diamond growth usually results in sparse, isolated diamond islands, with densities of $10^4-10^5$cm$^{-2}$  for growth on silicon. Hence, a seeding technique is needed to realise {coalesced} diamond thin films on  {foreign substrates}. {With} a {similar large difference in} surface energy\cite{northrup1996} {between diamond and GaN with a value} of $\sim 2$ J/m$^2$, a seeding technique is {also then required}  for diamond thin film growth {on this compound}. Historically, a variety of seeding techniques have been used to grow diamond on non-diamond substrates {from mechanical abrasion with diamond grit to bias enhanced nucleation\cite{williams2011rev}}. In this work we have determined the zeta potential of both Ga and N faced GaN. The result has been used to select the most suitable diamond particle for seeding the GaN wafers leading to self assembly of monolayer of diamond nanoparticles. The seeded wafers have also been imaged using atomic force microscopy to detect the effectiveness of the seeding. Finally, the wafers with both types of seeds were exposed to diamond growth conditions to grow this diamond films on GaN.

\section{Experiment}
The results presented in this article consider Ga and N-face GaN. The two different orientations were fabricated by using samples grown by the complementary techniques of metal organic chemical vapour deposition (MOCVD) and molecular beam epitaxy (MBE) which are known to generate Ga-face and N- face layers  respectively for standard growth conditions. Both the samples were grown on to 2-inch sapphire substrates with the Ga-face samples produced in an Thomas Swan close-coupled showerhead 6x2" system, using trimethygallium (TMGa), and Ammonia (NH$_3$) as sources. The details of the growth method are given in Datta et al.\cite{db2009}. For the N-polar samples growth was carried out in a VG V80H MBE system with an SVTa nitrogen plasma source.  The wafers were first degassed at 450 $^o$ eurotherm for 24 hours prior to entering the growth system to reduce contamination.  Sample growths were carried out using 1 sccm of nitrogen gas flow, and a plasma power of 350W, with 0W reflected power.  A 20 minute surface nitridation was carried out on the substrate surface at 750 $^o$C $\pm$ 30 $^o$C (measured by optical pyrometry, using a 50 nm layer of IR-absorbing molybdenum deposited on the back of the wafers prior to growth). The GaN was then deposited with a Ga flux of 17 nA and a growth time of 1hr 30 mins to give a layer thickness of approximately 600nm.

Streaming potential for both types of wafers were measured using {a} Surpass$^{\mbox{\tiny TM}}$ 3 {electrokinetic analyser}. The streaming potential is determined by measuring the change in potential or current between two Ag/AgCl electrode{s at either end of a streaming channel} as a function of electrolyte pressure. The electrolyte flowing through the narrow channel shears the counter-ions from the charged surfaces in contact with the liquid. The flowing charge will create a streaming current across two electrodes positioned tangential to the flow. With the flow of counter ions dependent on the electric double layer of the surfaces, measurement of the voltage then allows determination of the zeta-potential.  Streaming potential measurements have {thus} been used to determine the zeta-potential of fibers\cite{jacob1985}. For measurement of streaming potential on flat surfaces {meanwhile} a setup was suggested by Van-Wagenen et al.\cite{van1980} {, and has} been successfully used by many authors for measurement on various flat surface geometries \cite{voigt1983, norde1990, scales1990, werner1996}. {The streaming channel was formed between two plates of  Ga or N faced wafers placed parallel to each other at a distance of between 90 and 110 $\mu$m and a 10$^{-3}$ molar solution of potassium chloride was used as {the} electrolyte and the pressure varied from 600 to 200 mbar.  The pH of the electrolyte was {then altered with}  0.1M HCl and 0.1M NaOH solutions using the inbuilt titrator in Surpass$^{\mbox{TM}}$ 3. 

{Seeding was carried out with}  mono-dispersed diamond/H$_2$O solutions in an ultrasonic bath{,} known to produce nucleation densities in excess of 10$^{11}$ cm$^{-2}$  {atop} silicon wafers\cite{williams2007seed}. Two types of solution were used with both the wafers,  one containing particle{s} which were hydrogen terminated and the other containing particles with oxygen termination.  {While, commercially} available nanodiamond {particles} are oxygen terminated\cite{mochalin2011}, the process for hydrogen\cite{williams2010hyd} termination of nanodiamond  {particles}  {is detailed} elsewhere. The oxygen and hydrogen terminated particles have negative and positive zeta potential in water respectively. {The particle size as indicated by dynamic light scattering was found to be approximately 7 nm for both of the seeding solutions used.}  To  {investigate} the seeding density of the samples atomic force microscopy (AFM) images were taken using a Veeco Dimension 3100 {equipped} with RTESP tip{s (Bruker-Nano supplied, 8 nm nominal radius)} in tapping mode. {Post AFM analysis was carried out with WSXM\cite{horcas2007} and Gwyddion\cite{necas2012} SPM analysis sofware.} Scanning electron microscopy (SEM) images were taken using a Raith e-line operating at 20kV and 10mm working distance.  

For the growth of diamond on GaN-on-sapphire the samples were diced into 15 $\times$ 15 mm pieces and seeded. The seeded wafers were {then placed} in a Seki Technotron AX6500X microwave chemical vapor deposition system for diamond growth. {Growth} was carried under 5\% CH$_4$/H$_2$ condition at 40 torr pressure and 3.5kW microwave power. After growth the  {samples were} cooled slowly in hydrogen atmosphere to prevent cracking of the sapphire substrate. The growth of diamond on GaN was carried out on both Ga and N-face of GaN.

\section{Results and Discussion}
\begin{figure}[t]
\includegraphics[width=0.4\textwidth]{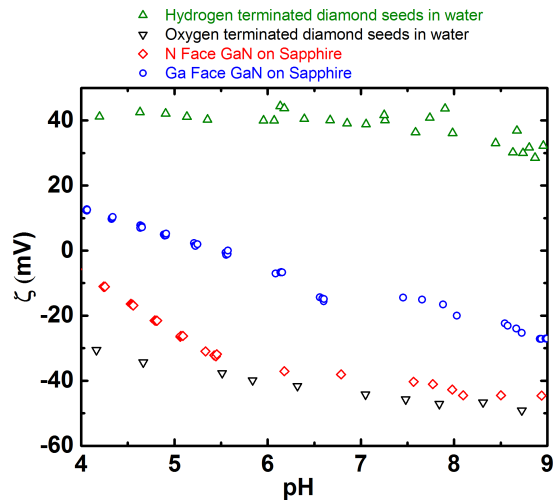}
\caption{Zeta potential versus pH curve for Ga and N-face GaN, Hydrogen and oxygen terminated diamond seeds in water are shown. The hydrogen terminated seeds show positive zeta potential in the pH rage 4-9, the Ga-face GaN has an isoelectric point at pH $\sim$5.5 and rest of the materials show varying degrees of negative zeta potential. The data for the seeds have been taken from Hees et al. \cite{hees2011}} \label{fig1}
\end{figure}
 {Figure 1 details} the zeta potential of Ga and N-face GaN surface as a function of pH. The zeta potential for N-face GaN is negative and there is moderate variation in the potential as the pH changes from 4 to 9. A similar trend is seen for Ga-face GaN which has an isoelectric point at pH $\sim$5.5 with negative zeta potential at higher pH values. {With water being the solvent for the diamond nano-particle seeding solutions, the main region} of interest for diamond growth are the values in the pH range 6 to 7. In this range N-face GaN shows higher negative zeta potential when compared to Ga-face GaN. This can be attributed to two different causes, i)  the presence of higher concentration of adsorbed oxygen on the N-face GaN\cite{hellman1996,zywietz1999}. and, ii) the Ga-face GaN surface is generally smoother than the N-face GaN surface\cite{tarsa1997}, as is evident from the AFM images in panels A and E in figure \ref{fig2} {and leading} to a minor inaccuracy in the calculation of the total surface area in contact with the streaming electrolyte.  Foster et al.\cite{foster2013} have shown that under aqueous conditions the N-face of GaN oxidizes more rapidly than that of Ga-face. The effect of oxidation of activated carbon on zeta potential was studied by Chingombe et al.\cite{chingombe2005} {, with the authors finding} that {the} higher the oxygen content on the surface of activated carbon the more negative was the zeta potential at a given pH value.  A similar trend can be seen for hydrogen and oxygen terminated nanodiamonds as well, the zeta potential vs pH for which have been plotted in figure \ref{fig1}. While the oxygen terminated seeds show negative zeta potential for the whole pH range, hydrogen terminated seeds have a positive zeta potential in the same range. For self assembly of a monolayer of nanodiamond on the GaN surface it is important to have the GaN and seed zeta potential of opposite polarity with highest possible difference. The results from the zeta potential study of the GaN surface and diamond seeds show that such conditions are met when the seeds are H-terminated and the pH of the solution is close to 8.  Another interesting feature to note in the zeta potential is the change in the potential as the pH is changed. Foster et al.\cite{foster2013} have shown that as the GaN surface is exposed to higher pH the oxygen content of the surface increases, which can lead to higher negative zeta potential. On the other hand exposure to aq-HCL leads to chlorine chemisorption which lowers oxide content\cite{uhlrich2008} on the surface resulting in lower negative zeta potential and eventually an isoelectric point on the Ga-face GaN. 

\begin{figure*}[t]
\includegraphics[width=0.6\textwidth]{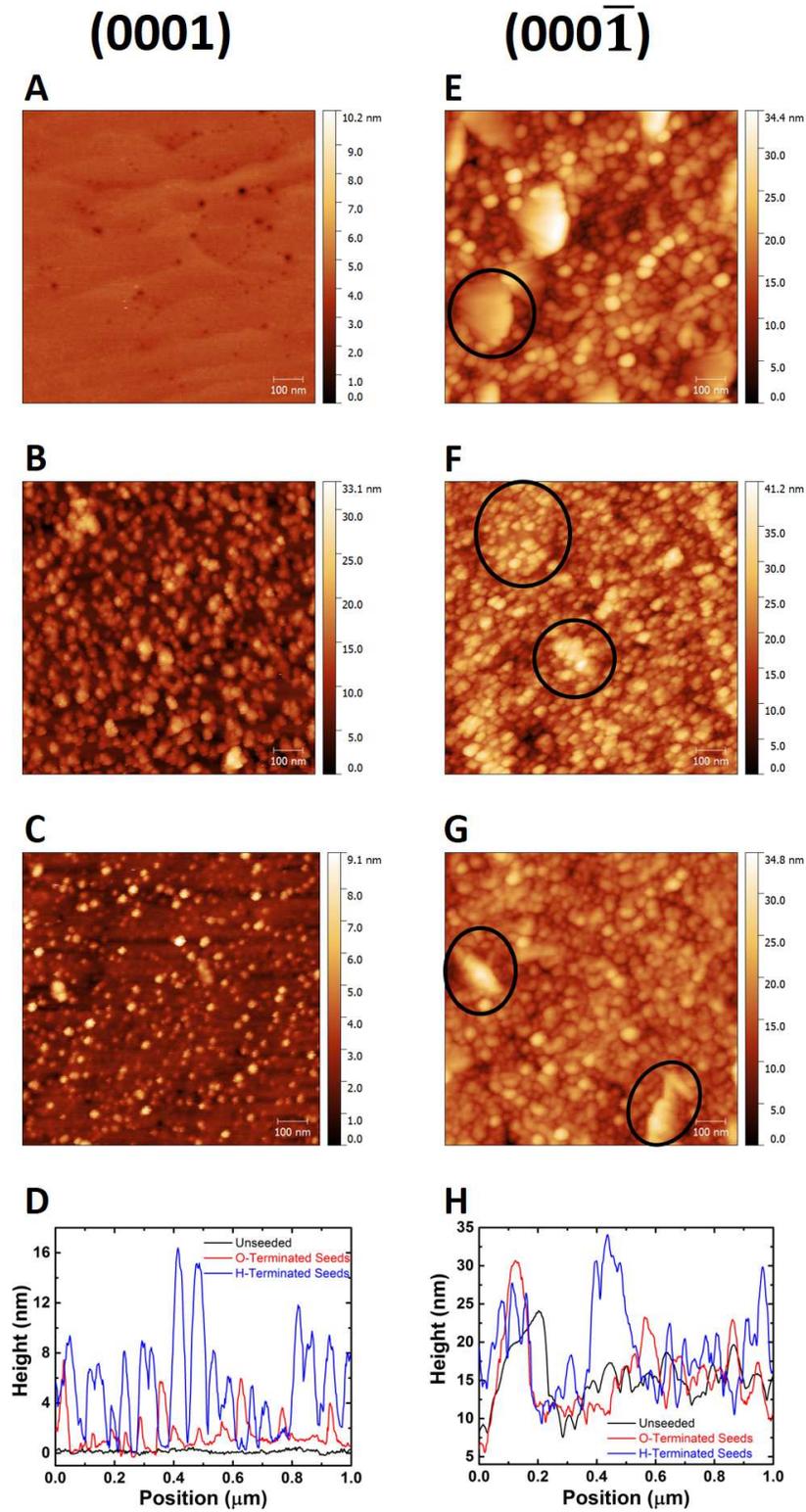}
\caption{AFM images of seeded and unseeded Ga and N-face GaN. The left hand side (Panels A-D) shows images from Ga-face GaN while the right hand side (Panels E-H) corresponds to N-face GaN. Panel A and E are for the respective unseeded GaN. Panels B and F shows the surface after seeding with H-terminated seeds. Panels C and G show the GaN surfaces after seeding with O-terminated seeds. Panels D and H are the line profiles of unseeded and seeded surfaces for both types of GaN} \label{fig2}
\end{figure*}

To compare the effectiveness of the seeding solutions  AFM {was performed} on seeded and unseeded GaN surfaces {with the results} shown in figure \ref{fig2}. Panels A and E of figure \ref{fig2} shows the AFM image of unseeded surfaces of Ga and N-face GaN. The second panels on both sides of the figure (Panels B and F) show the sample surface after being seeded by H-terminated diamond seeds. While the seeds are clearly visible on the smooth Ga-face GaN surface the same is not so clearly obvious for N-face GaN. The reason is the surface roughness of this surface. However, {focusing} on the large{r} crystallites on the surface in both figure{s} (marked on the images) {it can be seen} that while in panel E the crystallites are completely exposed{,} similar {crystallites} in panel F {show evidence of being covered by a conformal layer of particles}. Comparing {these} small particles with the ones {discernable} in panel B it is quite clear that there is a very high density of nanodiamond particles also on this surface. {In addition  pieces of both the Ga and N faced wafers were seeded with the O-terminated seeds.} The AFM images of such samples are shown in panels C and G of figure \ref{fig2}. On the Ga-face surface a direct comparison between panels B and C shows that the O-treated seeds produce a less densely seeded surface than the H-terminated seeds. The N-face surface requires a similar approach as taken for the H-terminated seeds considering the AFM images on larger crystallites present on the surface. The large crystallites on panel G show clear faceting and the usual granularity introduced by seeds is missing. Finally to compare more closely the three surfaces (unseeded and H and O-terminated seeded) for both types(Ga and N-Face) of material we have taken a line profile from the AFM data. The plots are shown in panels D and H. The Ga-face surface (panel D) is very easy to interpret. The unseeded line profile is presented in black and shows the presence of no peaks or valleys. The O and H terminated seeded surfaces are presented in red and blue respectively. While the blue curve shows large number of peaks ranging between 4-16 nm, the number of peaks for the red curve is much lower, showing a reduced seeding density. For comparing the surfaces of N-face GaN we have purposely taken the line profiles in such a way that at least one crystallite falls in the profile for each surface. As before, the unseeded surface is plotted in black, O-terminated seeded surface in red and H-terminated surface in blue. For all three curves there is a peak between positions 0 and 0.2$\mu$m. While for unseeded and O-terminated nanodiamond treated surfaces we do not see any sub-features in the peak, clear sub-features can be seen on the H-terminated nanodiamond treated surface. This clearly shows that while H-terminated seeds produce a high density of seeds on the surface the O-terminated seeds fail to do so, in-line with the zeta potential measurements of the {substrates}.  {From} figure \ref{fig2} it is {therefore} clear that H-terminated nanodiamond seed solution is the best option for creating a high density seeded surface in both types of GaN.

\begin{figure}[h]
\includegraphics[width=0.4\textwidth]{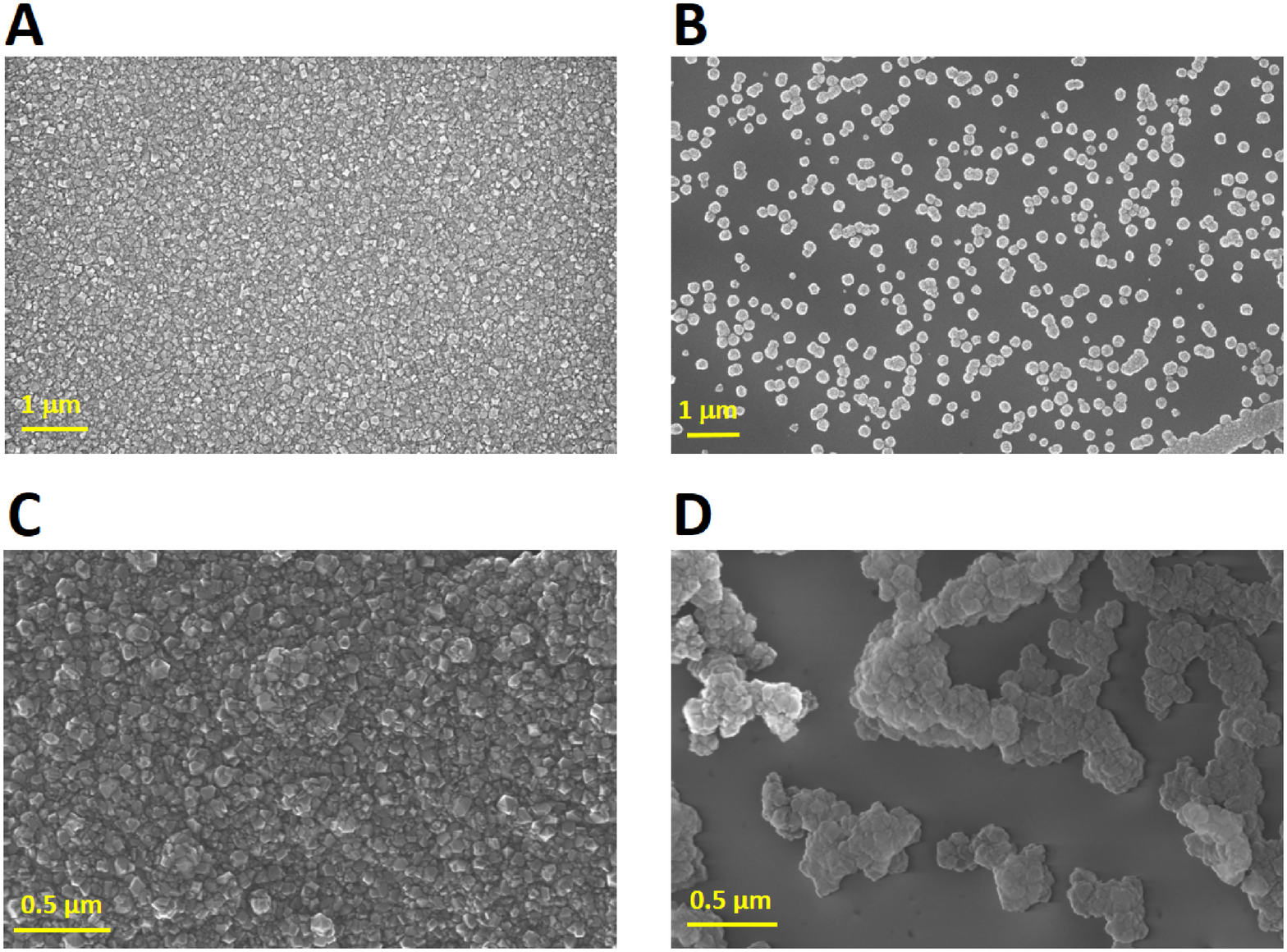}
\caption{SEM image of diamond thin films grown on GaN. Panel A shows the growth on surface treated with H-terminated diamond seeds while Panel B shows the sample treated with O-terminated seed solution on the Ga-Face of GaN. Panel C is the image for diamond growth on H-terminated diamond seeds treated N-face of GaN. Panel D shows growth for O-terminated seed treated N-face GaN showing non-coalesced growth.} \label{fig3}
\end{figure}
Finally, to test the effectiveness of seeding with H-terminated seeds, diamond thin films were grown on seeded (H-terminated and O-terminated) GaN. The SEM images of the grown films were taken and the results are shown in figure \ref{fig3}. Panel A and C in the figure shows the growth on GaN treated with H-terminated diamond seed solution for Ga and N-face respectively. We can see a complete pinhole free film indicating a seeding density greater than 10$^{12}$cm$^{-2}$. Panel B and D in the same figure show the samples treated with O-terminated diamond seed solution for Ga and N-face respectively. In this case only isolated crystals can be seen for Ga-face. Similarly, for N-face we see non-coalesced growth. The growth that is seen on the N-face can be attributed to the presence of initial surface roughness which can generate spontaneous nucleation sites\cite{spitsyn1981, dennig1991} but with low site density. So, from these figures it is clear that O-terminated seeds lead to very low seeding density.

\section{Conclusion}
The zeta potentials of both Ga and N-face GaN grown on sapphire as a function of pH were measured. Both surfaces show negative zeta potential between pH 5.5 and 9. The N-face GaN has a higher negative zeta potential due to larger concentration of adsorbed oxygen on the surface. The Ga-face shows an isoelectric point at $\sim$ pH 5.5. The results from the zeta potential study confirms the need for H-terminated seed solution at pH 8 to drive the self assembly of a nanodiamond monolayer on the GaN surface. AFM studies on seeded Ga- and N- face GaN confirmed the high seeding density when the surfaces were treated with H-terminated seeds. Subsequent growth of thin diamond films on the seeded wafers resulted in a pinhole free diamond layer on the surface treated with H-terminated seeds for both orientations of GaN. This technique of growing diamond directly on GaN removes the need for a seeding layer like silicon nitride with low thermal conductivity.

\section{Acknowledgment}
This project has been supported by Engineering and Physical Sciences Research Council under programme Grant GaN-DaME (EP/P00945X/1). DJW would like to acknowledge the support of EPSRC Manufacturing fellowship EP/N01202X/1. LEG would like to acknowledge the support of Royal Society starting grant "SCOPE". The metadata for the results presented in this paper can be found here().

\bibliography{ref}

\end{document}